\documentclass[prl,aps,amssymb,showpacs,twocolumn]{revtex4}
\usepackage{amsmath}
\usepackage{graphicx}

\begin{document}

\title{How to measure the spreading width for decay of superdeformed nuclei}

\author{D. M. Cardamone, C. A. Stafford, B. R. Barrett}
\affiliation{Physics Department, P. O. Box 210081, University of Arizona, 
Tucson, AZ 85721}

\date{April 2, 2003} 

\begin{abstract}
A new expression for the branching ratio for the decay via the $E1$ process
in the normal-deformed band of superdeformed nuclei is given within a simple
two-level model.  Using this expression, 
the spreading or tunneling width $\Gamma^\downarrow$
for superdeformed decay can be expressed
entirely in terms of experimentally known quantities. 
We show how to determine the tunneling matrix element $V$ from the measured
value of $\Gamma^\downarrow$ and a statistical model of the energy levels.
The accuracy of the two-level approximation is verified
by considering the effects of the other normal-deformed states.
\end{abstract}


\maketitle

Since the first discovery of superdeformation in $^{152}$Dy \cite{Twin86}, one
of the principal challenges has been to develop a consistent theory regarding
the decay-out mechanism of the superdeformed (SD) rotational band into the
normal-deformed (ND) band.  Although much experimental progress has been 
made since this first discovery, {\it e.g.,} Lauritsen {\it et al.} 
\cite{Dy02} and references therein, no consistent theory has been achieved
for this decay-out process, and, in fact, considerable confusion still
exists regarding the application of the current theoretical interpretations.  
The purpose of this Letter is to report on new theoretical developments,
which permit a direct determination of the spreading (or tunneling) width
for decay out of the SD band in nuclei in terms of
experimental quantities and, thereby, to obtain the magnitude of the
tunneling matrix element.

\begin{figure}
\includegraphics[width=.75\columnwidth, keepaspectratio=true]{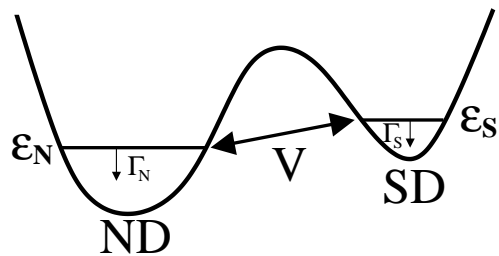}
\caption{Schematic diagram of the two-level problem. $V$ is the
  tunneling matrix element connecting the two states. Electromagnetic
  decay within each band gives the states their finite widths
  $\Gamma_N$ and $\Gamma_S$. $\varepsilon_S$ and $\varepsilon_N$ are
  the energies of the two levels in the absence of $V$.}
\label{2well}
\end{figure}

In an earlier publication \cite{SB99}, 
two of us presented a simple two-level model to 
explain the decay out of the SD band in nuclei.  Employing a retarded Green's
function approach, we obtained an exact solution for the branching ratio,
$F_N$, for decay via the $E1$ process in the ND band, in terms of the decay 
widths $\Gamma_S$ and $\Gamma_N$ in the SD and ND potential wells, respectively;
$V$, the tunneling matrix element connecting the SD state with the ND state,
{\it i.e.,} the state with which it mixes most strongly; and
$\Delta=\varepsilon_N-\varepsilon_S$, the
difference between the unperturbed energies of these two states (see Fig.\ \ref{2well} for a graphical 
representation of these quantities).  Our result yielded
\begin{equation}
F_N = \frac{(1+\Gamma_N/\Gamma_S)V^2}{\Delta^2 + \bar{\Gamma}^2
(1+4V^2/\Gamma_N\Gamma_S)},
\label{fN.result}
\end{equation}
where $\bar{\Gamma}=(\Gamma_S+\Gamma_N)/2$ and $V$ is taken to
be positive definite without loss of generality.

We have recently observed that Eq.\ (\ref{fN.result}) can also be rewritten
in the form
\begin{equation} 
F_N = \frac{\Gamma_N\Gamma^{\downarrow}/(\Gamma_N + \Gamma^{\downarrow})}
{\Gamma_S + \Gamma_N\Gamma^{\downarrow}/(\Gamma_N + \Gamma^{\downarrow})},
\label{fN.new} 
\end{equation} 
where 
\begin{equation}
\Gamma^{\downarrow} = \frac{\displaystyle 2\bar{\Gamma} V^2}{
\displaystyle \Delta^2 +\bar{\Gamma}^2}
\label{gamma.right}
\end{equation}
is the {\bf {\it correct}} expression from Fermi's golden rule for the spreading
(or tunneling) width \cite{SB99}. 
Eq.\ (\ref{fN.new}) clearly shows that
decay into the ND band is a two-step process, and that $\Gamma^\downarrow$ is
a real, physical rate, not a mere theoretical construct.
Note that
\begin{equation}
\Gamma^{\downarrow} \ne 2\pi\langle V^2\rangle/D_N,
\label{gamma.wrong}
\end{equation}
as was employed by Vigezzi {\it et al.} \cite{vigezzi}.  In 
Eq.\ (\ref{gamma.wrong}), $\langle V^2\rangle$ is the mean square of the 
coupling matrix elements connecting the SD and ND states and $D_N$ is 
the average spacing of the ND states.  In fact, Eq.\ (\ref{gamma.wrong}) 
gives the {\bf {\it average}} spreading width over a flat distribution
in $\Delta$ \cite{SB99}, 
which can deviate drastically from the exact value given by
Eq.\ (\ref{gamma.right}).

Importantly,
$\Gamma^{\downarrow}$, as given by Eq.\  
(\ref{gamma.right}),
is a  measureable quantity, which can be determined from Eq.\ 
(\ref{fN.new}):
\begin{equation} 
\Gamma^{\downarrow} = \frac{F_N\Gamma_N\Gamma_S} 
{\Gamma_N - F_N(\Gamma_N + \Gamma_S)}.
\label{gamma.solve} 
\end{equation} 
$\Gamma^\downarrow$ is {\it not} a free parameter as was stated in 
Ref.\ \onlinecite{Dy02}. 

We have used Eq.\ (\ref{gamma.solve}) to determine $\Gamma^{\downarrow}$
from the values of $F_N$, $\Gamma_S$, and $\Gamma_N$ given in Table I of
Ref.\ \onlinecite{Dy02}, and obtain values differing radically from those 
listed in Ref.\ \onlinecite{Dy02}, by three to five orders of magnitude
(see our Table \ref{table:widths}).  
This is a direct contradiction of the statement in 
Ref.\ \onlinecite{Dy02} that ``For cases encountered in experiments, 
their results (\emph{i.e.}, the results of our Ref.\ \onlinecite{SB99}) agree with those obtained with the
model of Vigezzi {\it et al.},'' {\it i.e.,} the model that was
employed in Ref.\ \onlinecite{Dy02}.

\begin{table}
\caption{Tunneling widths $\Gamma^{\downarrow}$ extracted from Eq.\ 
(\ref{gamma.solve}) compared with the results given in Table I
of Ref.\ \onlinecite{Dy02} (denoted by ${\Gamma^\downarrow}^{(2)}$).
$F_N \equiv P_{out}$, 
$\Gamma_S, \Gamma_N,$ and $D_N$ are the same as those in Table I of 
Ref.\ \onlinecite{Dy02}.  The spin values of the decaying states 
are given in parentheses. 
%$\Gamma^\downarrow$ is calculated using our
%Eq.\ (\ref{gamma.solve}), while ${\Gamma^\downarrow}^{(2)}$ is the
%result from Ref.\ \onlinecite{Dy02}.
}
\label{table:widths}
\begin{ruledtabular}
\begin{tabular}{ccccccc}
Nucleus & $F_N$ & $\Gamma_S$  & $\Gamma_N$   
& $D_N$  & $\Gamma^\downarrow$  
& ${\Gamma^\downarrow}^{(2)}$\\
  & $\equiv P_{out}$ & (meV) & (meV) & (eV) & (meV) & (meV) \\
\colrule
$^{152}$Dy(28) & 0.40 & 10.0  &  17 & 220 &  11   & 41,000 \\
$^{152}$Dy(26) & 0.81 &  7.0  &  17 & 194 & -40   & 220,000 \\
$^{194}$Hg(12) & 0.40 & 0.108 &  21 & 344 & 0.072 & 560 \\
$^{194}$Hg(10) & 0.97 & 0.046 &  20 & 493 &  1.6  & 37,000 \\
\end{tabular}
\end{ruledtabular}
\end{table}

It should be noted that we obtain a negative value for 
$\Gamma^{\downarrow}$ in $^{152}$Dy with $I = 26$.  This is, of
course, physically impossible: $\Gamma^{\downarrow}$ is
a positive definite quantity. Eq.\ (\ref{gamma.solve}) therefore requires
\begin{equation}  
F_N < \frac{\Gamma_N}{\Gamma_N + \Gamma_S}.
\label{fn.limit}  
\end{equation}   
Experimental results in which 
the inequality (\ref{fn.limit}) is violated may indicate that the
experimentally measured value of $F_N$ is too large or that
the SD state is mixing with a second ND level, since our
result is obtained for mixing with only one ND state.  At the
present time, however, uncertainties in the known values of
$\Gamma_N$ and $\Gamma_S$, which are of the order of 100\%
for $\Gamma_N$ and of the order of 10\% for $\Gamma_S$, or 
more, mean that $\Gamma^{\downarrow}$ cannot be meaningfully
determined in cases such as $^{152}$Dy(I=26). 

In order to determine the tunneling matrix element $V$ from  
$\Gamma^{\downarrow}$ via Eq.\ (\ref{gamma.right}), 
we must know $\Delta$, which generally is not 
experimentally known.  We therefore compute the expected value 
for $\Delta$ based on the assumption that the states to which the SD 
state decays in the ND well are distributed according to a  
Gaussian orthogonal ensemble (GOE). In the GOE, the probable spacing
between levels is given by the distribution \cite{mehta}
\begin{equation}
P(s) =  \frac{\pi}{2} s e^{-{\pi}s^{2}/4},
\label{prob.s}
\end{equation}
where $s$ is the spacing in units of $D_N$.

\begin{figure}
\includegraphics[width=\columnwidth, height=.5\columnwidth]{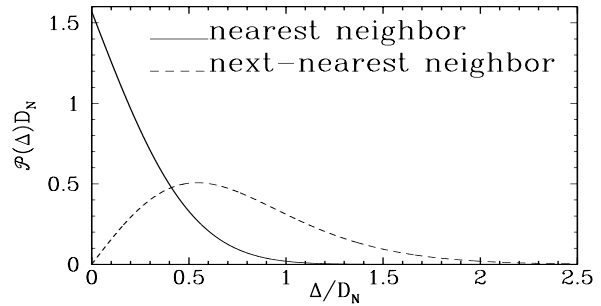}
\caption{Probability distributions for the two ND levels 
bracketing the SD level of interest,
from Eqs.\ (\ref{prob.delta}) and (\ref{prob.delta2}).
Note that the mean nearest-neighbor spacing is $\langle |\Delta_1|\rangle=
D_N/4$, while the mean spacing of the next-nearest neighbor is
$\langle |\Delta_2|\rangle= 3D_N/4$.}
\label{probs}
\end{figure}

In the absence of tunneling ($V=0$),
the energy spectra of the ND and SD wells are uncorrelated. Given a
spacing $sD_N$ between the nearest ND levels above and below the
decaying SD
level, then, $\Delta$ is given by the rectangular probability distribution
\begin{equation}
{\cal P}_{s}(\Delta) = \frac{1}{sD_N}{\Theta}\left(\frac{s}{2} -
\frac{|\Delta|}{D_N}\right).
\label{probdel.s}
\end{equation}
Here $\Theta$ is the Heaviside step function, which simply ensures
that $\Delta$ is the nearest neighbor.
The probability distribution for $\Delta$ is then
\begin{equation}
{\cal P}(\Delta) = \int_{0}^{\infty} {\cal P}_{s}\left(\Delta\right)
P(s) ds
=\frac{\pi}{2D_N}\mathrm{erfc}\left(\sqrt{\pi}\frac{|\Delta|}{D_N}\right).
\label{prob.delta}
\end{equation}
Figure \ref{probs} shows ${\cal P}(\Delta)D_N$ plotted as a function of
$\Delta/D_N$. From Eq.\ (\ref{prob.delta}), it is easy to compute the
average detuning $\langle|\Delta|\rangle=D_N/4$.

Our ultimate goal is to find the probability distribution ${\cal P}(V)$ for
given values of $\Gamma^{\downarrow}, \bar{\Gamma}$ and $D_N$, which in
general is given by ${\cal P}(V)=2{\cal P}\left(\Delta\right)\left|\frac{d\Delta}{dV}\right|$.
From Eq.\ (\ref{gamma.right}) we can obtain $|\Delta|$ as a function of
$V$,
\begin{equation} 
|\Delta|  =  \sqrt{\frac{2\bar{\Gamma}}{\Gamma^{\downarrow}}
\left(V^{2} - \frac{\Gamma^{\downarrow}\bar{\Gamma}}{2}\right)}.
\label{delta} 
\end{equation} 
$V$ obviously has a lower bound of
$V_{min}=\sqrt{\frac{1}{2}\Gamma^\downarrow\bar{\Gamma}}$ due to the requirement that
$|\Delta|$ be real.

Computing ${\cal P}(V)$ for the allowed region, we find
\begin{equation}
{\cal
  P}(V\geq V_{min})=\frac{2\bar{\Gamma}V\pi}{\Gamma^\downarrow|\Delta|D_N}
\mathrm{erfc}\left(\sqrt{\pi}\frac{|\Delta|}{D_N}\right),
\label{prob.v}
\end{equation}
where $|\Delta|$ is given by Eq.\ (\ref{delta}). The
average value of $V$ is
\begin{equation}
\langle V\rangle=\sqrt{\frac{\Gamma^\downarrow}{2\bar{\Gamma}}}
\left[\frac{D_N}{4}+{\cal O}\left(\frac{\bar{\Gamma}^2}{D_N}\right)\right].
\label{ave.v}
\end{equation}

Earlier attempts to consider a statistical theory of SD decay-out
\cite{vigezzi, WvBB98} have
been handicapped by their focus on average values of
$\Gamma^\downarrow$ and $F_N$. As was shown in Ref.\
\onlinecite{SB99}, the statistical fluctuations of these quantities
are much greater than their averages, making such computations almost
meaningless. This is not the case with the distribution of $V$ given
by Eq.\ (\ref{prob.v}),
however, for which the fluctuations are of order $\langle
V\rangle$. Furthermore, $\Gamma^\downarrow$ and $F_N$ are
experimentally fixed, whereas the uncertainty in $V$ is due to
lack of knowledge of the unperturbed energies of the ND and SD
states. Eq.\ (\ref{prob.v}) therefore represents a
central result, since it is the maximal information we can have
about $V$ without prior knowledge of the shape of the potential.

Thus far we have treated only a two-level model of
the superdeformed decay-out process. It is reasonable to ask what
effect the inclusion of more ND states could have. As a first step, we
include the next-nearest neighbor state in the ND well.

We can again
turn to
the assumption of the GOE in order to find typical values
for $\Delta_1$ and $\Delta_2$, the energies of the nearest and next-nearest 
neighbor states, respectively. ${\cal
  P}(\Delta_1)$ is simply the ${\cal P}(\Delta)$ given by Eq.\
(\ref{prob.delta}). The calculation of
${\cal P}(\Delta_2)$ is similar to that for ${\cal P}(\Delta_1)$, but we must now concern ourselves with the question
of whether $\Delta_1$ and $\Delta_2$ have the same or opposite
signs.

In all cases of physical interest, we find that for a
given spacing $s$, the
correction to the two-state result necessitated by the inclusion of
a third level is larger when the two ND levels bracket the SD level. 
This is quite natural, since Eq.\ (\ref{prob.s})
requires that the next-nearest ND level in this
configuration lie on average 40\% closer  
to the decaying SD state than
it would if the nearest two ND levels lay on the same side of the SD level.
Consideration of the
case in which $\Delta_1$ and $\Delta_2$ have the same sign, then, only
decreases the necessary correction. Since our goal is to set a reasonable upper
bound on this quantity, we assume the ND levels lie on opposite sides of
the SD level. Decays for which this is not true will in general
conform to the two-state approximation with even greater accuracy.

Based on this assumption, we can construct
a density function for $\Delta_2$ similar to Eq.\ (\ref{probdel.s}),
\begin{equation}
{\cal P}_s(\Delta_2)=\frac{1}{sD_N}\Theta\left(\frac{|\Delta_2|}{D_N}-\frac{s}{2}\right)\Theta\left(s-\frac{|\Delta_2|}{D_N}\right).
\end{equation}
Together with Eq.\ (\ref{prob.s}), this yields a distribution for $\Delta_2$
\begin{equation}
\label{prob.delta2}
{\cal P}(\Delta_2)=\frac{\pi}{2D_N}\left[\mathrm{erf}(\sqrt{\pi}|\Delta_2|)
  - \mathrm{erf}(\frac{\sqrt{\pi}}{2}|\Delta_2|)\right].
\end{equation}
This expression for ${\cal P}(\Delta_2)$ is illustrated by
Fig.\ \ref{probs}. Its average detuning is
$\langle|\Delta_2|\rangle=3D_N/4$.

Having computed the average values of $\Delta_1$, $\Delta_2$, and $V$,
we are now in a position to begin
to see the effect of a second ND level. In general, Eq.\
(\ref{fN.result}) would suggest that the contribution to the total
branching ratio of a second level is substantially less than that of
the 
nearest neighbor. Since Eq.\ (\ref{fN.result}) was derived in the
context of only one ND level, however, we ought to seek a more
rigorous theory for the three-state branching ratio. In particular, we
should expect that effects such as competition and interference will
play a role in the exact result.

The Hamiltonian for the three-state system can be taken to be the sum
of two parts, $H_0$, which represents the independant SD and ND wells,
and $\hat{V}$, which mixes the states of the two wells. $H_0$ can be written
\begin{equation}
H_0=\sum_i\varepsilon_ic_i^\dagger c_i+H_{EM},
\end{equation}
where the sum on $i$ runs over S, N1, and N2, $c_i$ is the anihilation
operator for state $i$, and $H_{EM}$ contains the coupling to the
electromagnetic field which gives the states their nonzero
widths. Since they occur by the same decay process, we assume the
widths of the ND states are equal, \emph{i.e.} $\Gamma_{N1}=\Gamma_{N2}=\Gamma_N$.

\begin{figure}
\includegraphics[width=\columnwidth,height=.8\columnwidth]{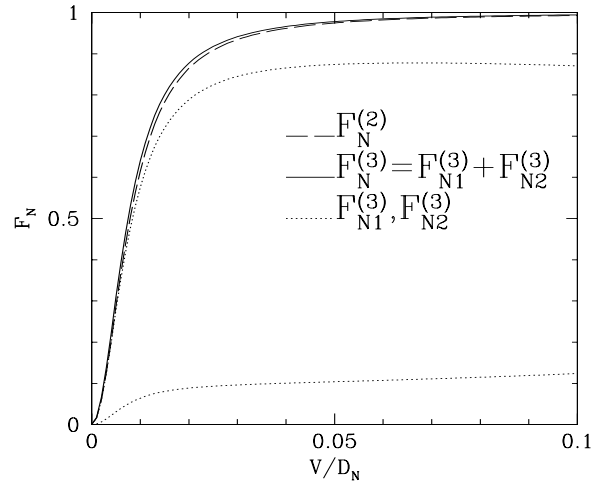}
\caption{Plot of ND branching ratios calculated in the two
  ($F_N^{(2)}$) and
  three-state ($F_N^{(3)}$) models with parameters relevant to the $A\approx 190$ mass
  region. The curves labeled $F_{N1}^{(3)}$ and
  $F_{N2}^{(3)}$ represent the branching ratios for the individual states of
  the three-level model, which sum to the total branching ratio. The
  energy levels were taken to lie at their mean detunings, and 
  constant values of $\Gamma_S/\Gamma_N=10^{-3}$ and
  $\bar{\Gamma}/D_N=10^{-4}$ were used. These orders of magnitude are consistent with
  Table\ \ref{table:widths}.}
\label{190graphs}
\end{figure}

Taking $\hat{V}$ as a perturbation, it is a trivial exercise to use Dyson's
Equation to construct the retarded Green's function of the
system. The result, exact to all orders in $\hat{V}$, is given in the
$|S\rangle$, $|N1\rangle$, $|N2\rangle$ basis by 
\begin{equation}
\label{G0}
G^{-1}(E)=\begin{pmatrix}
E+i\frac{\Gamma_S}{2}&-V_1&-V_2\\-V_1&E-\Delta_1+i\frac{\Gamma_N}{2}&0\\
-V_2&0&E-\Delta_2+i\frac{\Gamma_N}{2}\end{pmatrix},
\end{equation}
where $V_1$ and $V_2$ may be chosen positive without loss of generality.
In the following, we assume further that $V_1=V_2$.

The branching ratios of the full three-state system can now be computed from Parseval's theorem
\begin{equation}
F_i=\Gamma_i\int_{-\infty}^\infty\frac{dE}{2\pi}|\langle S|G(E)|i\rangle|^2,
\end{equation}
where $i=S,N1,N2$.
These integrals can be done analytically by Cauchy
integration, but the results are algebraically complicated. It is sufficient
for our purposes to compute them numerically. 

\begin{figure}
\includegraphics[width=\columnwidth,height=.8\columnwidth]{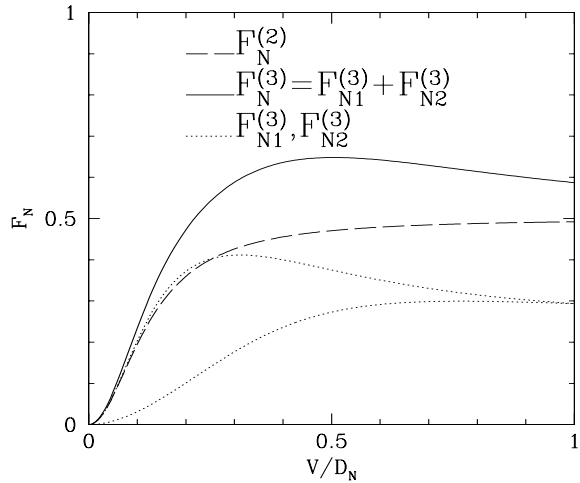}
\caption{Plot of ND branching ratios calculated in the two-
  and
  three-state models with parameters relevant to the $A\approx 150$
  mass region. The energy levels were taken to lie at their mean
  detunings, and
  constant values of $\Gamma_S=\Gamma_N$ and
  $\bar{\Gamma}/D_N=10^{-4}$ were used. These orders of magnitude are consistent with
  Table\ \ref{table:widths}. The notation for branching ratios
  used here
  is defined in the caption of Fig.\ \ref{190graphs}.}
\label{150graphs}
\end{figure}

With $\Delta_1$ and $\Delta_2$ determined by their probability
distributions, the only remaining parameters in the three-state
problem are $\Gamma_S$, $\Gamma_N$, $V$, and $D_N$. In all cases of physical
interest, however, $\Gamma_S,\Gamma_N\ll V,D_N$, so we can restrict the
relevant parameter space by varying only the combinations of
parameters $\Gamma_S/\Gamma_N$ and $V/D_N$ separately. The
corrections to the branching ratios required by such a restriction are
of order $\Gamma_N/V$.

Furthermore, the order of $\Gamma_S/\Gamma_N$ is determined by the
mass region of the nucleus (see Table\ \ref{table:widths}). In the
results that follow, variation of this parameter over reasonable
values does not significantly impact the necessary correction to the
two-state result.

Figures \ref{190graphs} and \ref{150graphs} show comparisons of the
two- and three-state branching ratios for the $A\approx 190$ and $A\approx 150$ mass regions,
respectively.
These figures demonstrate that in cases of physical interest, the
correction to the two-state system due to the presence of a third
level is relatively small. In the $A\approx 190$ region, in particular, 
we find that $.9\lesssim F_N^{(2)}/F_N^{(3)}\leq 1$, where the
superscripts on the branching ratios indicate the number of states
included in their calculation. In the
$A\approx 150$ region, we find that $0.7 <F_N^{(2)}/F_N^{(3)}< 1$. 
The increased importance of additional levels in the $A \approx 150$ 
region arises because the typical tunneling matrix elements at decay out
are significantly larger, which follows from Eq.\ (\ref{fN.result}) and
the relative sizes of $\Gamma_S$ and $\Gamma_N$ in the two mass regions.
Constructive interference between the two ND levels,
plainly visible in Fig.\ \ref{150graphs} where $F_N^{(3)}$ exhibits a
distinct maximum for $V\sim D_N/2$, can also enhance the importance of 
additional ND levels.

In the $190$ mass region, however, quantum interference effects are seen
to be negligable, and this approach can be extended to set an approximate upper bound on the
total error incurred by ignoring \emph{all} of the ND states except the
nearest neighbor. It is clear from Eq.\ (\ref{fN.result}) that, neglecting quantum interference
effects, a level's branching ratio is approximately
proportional to $\Delta^{-2}$. Assuming that each level in the
infinite ND band lies at its average detuning, we can write the total
$n$-level branching ratio as
\begin{equation}
F_N^{(n)}\cong\sum_{k=1}^{n-1}\frac{F_N^{(2)}(D_N/4)^2}{(\Delta_k)^2}=\sum_{k=1}^{n-1}\frac{F_N^{(2)}(D_N/4)^2}{\left[(2k-1)\frac{D_N}{4}\right]^2}.
\label{FNn}
\end{equation}
We thus have the result
\begin{equation}
F_N^{(\infty)}\cong\sum_{k=1}^{\infty}\frac{F_N^{(2)}}{(2k-1)^2}=
\frac{\pi^2}{8}F_N^{(2)}.
\label{FNinf}
\end{equation}
In the 190 mass region, then, the expected correction is no more than
about 23\% of $F_N^{(2)}$.

The three-state results, together with the arguments of
Eqs.\ (\ref{FNn}) and (\ref{FNinf}), demonstrate that the two-state model 
is sufficient to describe the dominant decay-out process of SD nuclei. 
Within the two-state model, we have shown that
the decay out of an SD level via $E1$
processes in the ND band is a two-step process, whose branching ratio 
(\ref{fN.new}) is
expressed in terms of three measureable rates, $\Gamma_S$, $\Gamma_N$,
and $\Gamma^\downarrow$.  We have also shown how to determine the 
tunneling matrix element $V$ [Eqs.\ (\ref{prob.v}) and (\ref{ave.v})]
from the measured value of $\Gamma^\downarrow$
and a statistical model of the ND band.
It is hoped that these results will help clarify the 
nature of the decay-out process in SD nuclei.

We thank T. L. Khoo and S. \AA berg for helpful discussions and
acknowledge support from NSF grant PHY-0210750.  B. R. B. 
acknowledges partial support from NSF grant No.\ PHY-0070858.
We also thank the Institute of Nuclear Theory at the University of
Washington for its hospitality and the Department of Energy for
partial support during the formulation and development of this
work.

\end{document}